\documentclass[%
 preprint,
 amsmath,amssymb,
 aps,
 prd,
]{revtex4-1}

\usepackage{graphicx}
\usepackage{dcolumn}
\usepackage{bm}
\usepackage{mciteplus}

\usepackage{mathrsfs}
\usepackage{slashed}
\usepackage{subfigure}
\usepackage{color}
\usepackage[normalem]{ulem}
\usepackage{enumerate}

\newcommand{\overbar}[1]{\mkern 2.5mu\overline{\mkern-2.5mu#1\mkern-2.5mu}\mkern 2.5mu}

\def\be{\begin{equation}}
\def\ee{\end{equation}}
\def\beq{\begin{equation}}
\def\eeq{\end{equation}}
\def\bc{\begin{center}}
\def\ec{\end{center}}
\def\bea{\begin{eqnarray}}
\def\eea{\end{eqnarray}}

\newcommand{\mean}[1]{\langle#1\rangle}

\newcommand{\mbD}{\mathbf{D}}
\newcommand{\mbTh}{\mathbf{\Theta}}

\newcommand{\lhood}{\mathcal{L}}
\newcommand{\ev}{\mathcal{Z}}

\newcommand{\ie}{\emph{i.e.}}
\newcommand{\eg}{\emph{e.g.}}
\newcommand{\df}{{\rm d}}
\newcommand{\qu}[1]{``#1''}
\newcommand{\eref}[1]{Eq.~(\ref{#1})}

\def\T2K{{\sc T2K }}

\def\MN{{\sc MultiNest}}
\def\HS{{\sc HiggsSignals 1.2.0}}

\newcommand{\refcite}[1]{Ref.~\cite{#1}}
\newcommand{\figref}[1]{Fig.~\ref{#1}}
\newcommand{\tabref}[1]{Tab.~\ref{#1}}
\newcommand{\secref}[1]{Sec.~\ref{#1}}

\def\SM{\text{SM}}
\newcommand{\nSM}{\overbar{\SM}} 

\begin{document}

\preprint{arXiv:1411.4876}

\title{Bayesian Model comparison of Higgs couplings}

\author{Johannes Bergstr\"{o}m}
	\email{bergstrom@ecm.ub.edu}
 \affiliation{Departament d'Estructura i Constituents de la Mat\`eria and Institut
  de Ciencies del Cosmos,\\ Universitat de Barcelona, Diagonal 647,
  E-08028 Barcelona, Spain}
\author{Stella Riad}%
 	\email{sriad@kth.se}
\affiliation{%
Department of Theoretical Physics, School of Engineering Sciences,\\ KTH Royal Institute of Technology  -- AlbaNova University Center,\\ Roslagstullsbacken 21, 106 91 Stockholm, Sweden
}%


\date{\today}

\begin{abstract}
We investigate the possibility of contributions from physics beyond the Standard Model (SM) to the Higgs couplings, in the light of the LHC data. The work is performed within an interim framework where the magnitude of the Higgs production and decay rates are rescaled though Higgs coupling scale factors. We perform Bayesian parameter inference on these scale factors, concluding that there is good compatibility with the SM. Furthermore, we carry out Bayesian model comparison on all models where any combination of scale factors can differ from their SM values and find that typically models with fewer free couplings are strongly favoured. We consider the evidence that each coupling individually equals the SM value, making the minimal assumptions on the other couplings. Finally, we make a comparison of the SM against a single \qu{not-SM} model, and find that there is moderate to strong evidence for the SM. 
\end{abstract}

%
\keywords{Statistical methods, Higgs physics}

\maketitle

\section{Introduction}
The discovery of a boson with a mass of approximately $125.5$ GeV was announced in July 2012 by the ATLAS and CMS experiments at the Large Hadron Collider (LHC) at CERN \cite{Aad:2012tfa,Chatrchyan:2012ufa}. This discovery is compatible with previous data from proton-antiproton collisions at $\sqrt{s}=1.96$ TeV at the Tevatron \cite{Aaltonen:2013kxa}. Using all of the available data, with a total luminosity of 25 ${\rm fb}^{-1}$ from the proton-proton collisions with energies of  $\sqrt{s} = 7$ and 8 TeV runs at the LHC, properties of the Higgs boson properties, such as spin, parity, mass, and the couplings to other Standard Model (SM) particles, has been further investigated \cite{ATLAS-CONF-2014-009,Chatrchyan:2013mxa,Chatrchyan:2013iaa, CMS-PAS-HIG-14-009}. So far, however, there are no indications of major deviations from the properties of the SM Higgs boson, and the boson does in fact seem to be a CP even scalar \cite{Aad:2013xqa,Chatrchyan:2012jja, Khachatryan:2014kca}. The discovery of the Higgs boson marks an important milestone in the history of particle physics, especially for our understanding of electroweak symmetry breaking and the generation of particle masses \cite{Higgs1964132,PhysRevLett.13.508,PhysRevLett.13.585,PhysRevLett.13.321}. 

The Higgs boson was discovered through its decays into gauge bosons, \ie, $WW$, $ZZ,$ and $\gamma\gamma$ pairs. However, using the full set of data from the LHC, there is now also evidence for decays into fermions, $b\bar{b}$ and $\tau^+\tau^-$ \cite{Chatrchyan:2014vua, ATLAS-CONF-2013-108}.

Even though the properties of the new boson so far are compatible with those of the SM Higgs, the possibility for new physics in the Higgs sector should be investigated. New physics can manifest itself in different ways, some of which can be detected since they would give rise to a rescaling of the magnitude, or change in the structure, of the Higgs boson couplings. Hence, a natural step forward in experimental Higgs physics is precision measurements of the Higgs boson couplings to fermions and gauge bosons. From existing data, there are bounds on the couplings of the boson. However, hadron colliders are in general not ideal for Higgs precision measurements and thus, in order to determine the couplings with significantly greater precision, the upgrade of the LHC to 14 TeV won't suffice and instead lepton colliders, such as a Higgs factory, are needed \cite{Peskin:2013xra}.

The status of the Higgs couplings as measured by the LHC can be studied by means of so-called \emph{Higgs coupling scale factors}, introduced by the LHC Higgs cross section working group as an interim treatment of the Higgs couplings \cite{LHCHiggsCrossSectionWorkingGroup:2012nn}. Coupling scale factors are introduced in order to rescale the magnitudes of the Higgs production and decay rates, which is especially useful since the experimental data from the collaborations are presented in terms of so-called \emph{global signal strengths}. Since the analysis of the data indicate that a CP-even scalar is preferred to a CP-odd one, we assume a single underlying CP-even scalar boson at a mass of about 125 GeV and furthermore, we assume a simplifying zero-width approximation. This so-called \emph{interim framework} has been used by the ATLAS and CMS collaborations as well as in several phenomenological studies \cite{ATLAS-CONF-2014-010,Belanger:2013xza, Giardino:2013bma, Ellis:2013lra, Baak:2014ora, Englert:2014uua, Bechtle:2014ewa, Ferreira:2014naa, Cacciapaglia:2014rla,Cheung:2013kla, Cheung:2014noa, CMS-PAS-HIG-14-009, ATLAS-CONF-2014-009}. In this work, we use the software \HS~for the implementation of the LHC data in the form of a $\chi^2$-function \cite{Bechtle:2008jh, Bechtle:2013xfa, Bechtle:2013wla}.

In the present work we apply Bayesian inference within the framework of coupling scale factors. We shall use Bayesian parameter inference to get a rough idea of how the parameters are constrained. However, since the most important question is rather that of which model best describes the data, we will focus on model comparison -- in particular of different models in which any combination of couplings differ from their SM values. This framework makes it possible to compare many models to each other at once, but the main advantage is that it is possible to obtain evidence \emph{in favour} of simpler models -- in the present case for models where the couplings are given by their SM values.

The paper is organized as follows. In Sec.~\ref{sec:higgs}, we give an introduction to Higgs physics and the concept of coupling scale factors. In Sec.~\ref{sec:statistics}, we discuss the Bayesian method used in the present work, especially model comparison in the context of Higgs couplings. In addition, we discuss the models used in the present work as well as the priors used. In Sec.~\ref{sec:results}, we discuss the results of parameter estimation, and the results concerning the different questions addressed using model comparison. Finally, in \secref{sec:conclusions}, we give a short summary and give our conclusions.

\section{Higgs physics}\label{sec:higgs}

Whether the discovered particle at 125.5 GeV actually is \emph{the} SM Higgs boson, or only a part of some bigger picture, is an important question which needs to be investigated. In general, additional degrees of freedom in the Higgs sector will influence the Higgs couplings to the SM particles as well as the loop-induced production and decay modes.

One common way to investigate the possibility of new physics in the Higgs sector is to study and compare specific renormalizable models for beyond the SM physics, such as Two-Higgs-Doublet Models \cite{Chang:2013ona,Celis:2013rcs,Celis2}, composite Higgs models \cite{PhysRevD.88.095006,Carena:2014ria}, a dilaton model \cite{Giardino:2013bma}, and supersymmetric models \cite{Craig:2013cxa,Blum:2012ii}. However, in these cases the comparisons are only made between these specific models and the SM, and obviously lack in generality. Another way is to consider the SM extended with effective operators, resulting from new physics above the TeV scale. Since this new physics is heavy by assumption it will give rise to modifications of the couplings, which are suppressed by the scale of new physics. These modifications are however not necessarily small in magnitude if the scale of new physics is low \cite{Grzadkowski:2010es,Contino:2013kra}. In both of the frameworks discussed above, some or all of the Higgs boson couplings will be altered, both the magnitude and in principle also the tensor structure of the couplings, even though these modifications are often heavily suppressed. Another approach is simply to not consider a physical and realistic model, but instead make a statistical analysis based on the \qu{naive} rescaling of the magnitude of the Higgs couplings. In such a framework it is only possible to investigate whether there are any significant deviations of the couplings from their SM values, without attempting to interpret the underlying physics. Thus the relevant result of the analysis is whether the couplings deviate from their SM value or not, rather than the exact value of the couplings. This treatment with coupling scale factors shall be considered here.

\subsection{Production modes}
Four production modes of the Higgs boson in the SM are significant at the LHC. The predominant production mode is the loop-induced gluon fusion $gg\rightarrow H$, with heavy quarks running in a triangular loop, with the main contribution coming from the top quark. Since this process is loop-induced it is of particular interest in searches for new physics. The subdominant processes are vector boson fusion, $qq'\rightarrow qq'H$, associated production with a vector boson, $q\bar{q}\rightarrow WH/ZH$, and the associated production with a top-quark pair, $q\bar{q}/gg \rightarrow t\bar{t}H$. We will use the notation where $l=e,\mu$ and $q$ stands for any quark. 

\subsection{Decay modes}
The Higgs boson can decay either to a fermion-antifermion pair or two gauge bosons. At present, the Higgs boson have been detected in five decay channels at the LHC, namely the $\gamma\gamma$, $ZZ^{(*)}$ (in turn followed by a decay to $4l,2l2\nu,2l2q,2l2\tau$), $WW^{(*)}$ (followed by decays to $l\nu l\nu$, $l\nu qq$), $b\bar{b}$, and $\tau^{+}\tau^{-}$, which then decay leptonically and hadronically. Since the Higgs' coupling to fermions is proportional to the fermion mass, the heaviest fermion mode, which is kinematically accessible, will have the largest partial decay width in the SM. Of the detected decay modes only $H\rightarrow\gamma\gamma$ is loop-induced, which are of particular interest for searches beyond the SM. In addition to the observed decay modes, the $H\rightarrow \mu^{+}\mu^{-}$ and $H\rightarrow Z\gamma$ channels have been investigated at the LHC. However, the Higgs boson has not been detected in either of them and there are therefore only (rather loose) upper bounds in these channels at present \cite{Aad:2014xva}.

\subsection{Definition of coupling scale factors} 
The LHC Higgs results are commonly presented in terms of global signal strengths, defined as
\begin{eqnarray}
\mu = \frac{\sigma(X)\cdot \text{BR}(H\rightarrow Y)}{\sigma(X)_{\rm SM}\cdot \text{BR}(H\rightarrow Y)_{\rm SM}},
\end{eqnarray}
where $\sigma(X)$ is the cross section for the production mode $X$ and ${\rm BR} (H\rightarrow Y)_\SM$ the branching ratio of the decay mode $Y$. In the case of a SM process the value of $\mu$ is naturally 1. In the SM the Higgs boson couples to the other particles with couplings $y^\SM_i$, where $i\in \{t,b,\tau, \mu, W,Z\}$. The couplings to the fermions are the Yukawa couplings
\be  y^\SM_f = \frac{m_f}{v}, \ee
where $m_f$ is the mass of the fermion, $f\in \{b,t,\tau,\mu\}$, and $v$ is the Higgs vacuum expectation value. The upper perturbative limit for these couplings is around $4\pi$. For the gauge couplings we have 
\begin{eqnarray}
y_W^\SM = \frac{2m_W^2}{v}, \hspace{2 mm} y_Z^\SM = \frac{m_Z^2}{v},
\end{eqnarray}
where $m_W$, $m_Z$ are the $W$ and $Z$ masses, respectively. Note that these couplings are dimensionful.

A simple extension of the SM can be made by rescaling the magnitude of the SM decay and production rates, which effectively leads to a rescaling of the Higgs couplings by so-called coupling scale factors, $\kappa_i$. For the processes which exist at tree-level in the SM, the couplings are rescaled as
\begin{equation}
y_{i} = \kappa_{i}\cdot y^{\rm SM}_{i}.
\end{equation} 
Naturally, the SM is recovered for $\kappa_i=1$. In addition, coupling scale factors can be introduced for the loop-induced processes. We introduce $\kappa_g$ and $\kappa_\gamma$ for the $gg\rightarrow H$ and $H\rightarrow \gamma\gamma$ respectively. In principle, a scale factor, $\kappa_{Z\gamma}$, could be introduced for a third loop-induced process $H\rightarrow Z\gamma$. However, since the sensitivity is nowhere close to the region of the SM prediction and since the inference for the other parameters will not be affected, we will not include this as a free parameter. The cross section of the process $ii\rightarrow H \rightarrow ff$ is then given by
\begin{eqnarray}
(\sigma \cdot {\rm BR})(ii \rightarrow H \rightarrow ff) = \sigma_{\rm{SM}}(ii \rightarrow H)\cdot {\rm BR}_\SM(H\rightarrow ff)\cdot\frac{\kappa_i^2\kappa_f^2}{\kappa_H^2},
\end{eqnarray}
where $\kappa_i$ and $\kappa_f$ corresponds to the initial and final states respectively and $\kappa_H$ is the scale factor for the total Higgs decay width.

The coupling scale factors $\kappa_g$ and $\kappa_\gamma$ can be considered either as functions of the other coupling scale factors or free parameters of the fit if new physics is allowed to participate in the loops. In the SM these scale factors have the values $\kappa_g = \kappa_\gamma=1$. However, in the case when only the tree-level scale factors are varied, the scale factors of the loop-induced processes will vary depending on the other scale factors. The effects of the rescaled tree-level couplings would have to be cancelled by some new physics, if these parameters were fixed to their SM values. If the scale couplings are free, new physics is allowed to propagate in the loop. 

Furthermore, the factor $\kappa_g$ can be defined in two different ways, either in terms of partial cross-sections or decay widths. In the present case we define the coupling scale factor $\kappa_g(\kappa_t,\kappa_b)$ using the cross sections, since gluon fusion is the more important process. Thus, the scale factor is given by
\begin{eqnarray}\label{eq:gluon}
\kappa^2_g(\kappa_b,\kappa_t)=\frac{\kappa_t^2\cdot \sigma_{ggH}^{tt}+\kappa_b^2\cdot\sigma_{ggH}^{bb}+\kappa_t\kappa_b\cdot\sigma_{ggH}^{tb}}{\sigma_{ggH}^{tt}+\sigma_{ggH}^{bb}+\sigma_{ggH}^{tb}}.
\end{eqnarray}
In terms of the other $\kappa$'s, $\kappa_\gamma$ is given by
\begin{equation}\label{eq:gamma}
\kappa^2_\gamma(\kappa_b,\kappa_t,\kappa_\tau,\kappa_W)=\frac{\sum_{ij}\kappa_i\kappa_j\cdot \Gamma_{\gamma\gamma}^{ij}}{\sum_{i,j}\Gamma_{\gamma\gamma}^{ij}},
\end{equation}
where $\Gamma_{\gamma\gamma}^{ij}$ are the partial decay widths and the pairs $(i,j)$ are given by $bb,\, tt,\,$ $\tau\tau,\, WW,\,bt,$ $b\tau,\, bW,\, t\tau,\, tW,\, \tau W$ \cite{LHCHiggsCrossSectionWorkingGroup:2012nn}.

In addition, the total Higgs width scales with a coupling scale factor, which is defined in terms of the other coupling scale factors as
\begin{equation}\label{eq:kappah}
\kappa_H^2=\sum_{X}\kappa_X^2\cdot {\rm BR_{SM}}(H\rightarrow X),
\end{equation} 
where the summation runs over all possible decay modes in the SM. This parametrization requires that the resonance width is small and therefore the zero-width approximation is assumed. In principle, new physics could contribute to the total Higgs width, which occurs if, for instance, the Higgs can decay to dark matter particles. In this case $\kappa_H$ should be a free parameter, see for example Ref.~\cite{Pospelov:2011yp}. For an extensive description of the concept of coupling scale factors, see Ref.~\cite{LHCHiggsCrossSectionWorkingGroup:2012nn}.

In the present work we shall focus on the coupling scale factors in two settings. First, the scale factors corresponding to the SM tree-level couplings (and which are currently constrained by LHC data), \ie, the Higgs couplings to $b\bar{b}$, $t\bar{t}$, $\tau^+\tau^-$, $\mu^+\mu^-$, $WW$, and $ZZ$, have the possibility to be varied. In the second case, the loop-induced processes have the potential to be scaled as well, through the variation of $\kappa_g$ and $\kappa_\gamma$. We shall not consider the total decay width to be a free parameter in the present case. Again, the information on effective scale couplings from LHC data were implemented using the \HS~software.

Note that the new particle is assumed to \qu{resemble} the SM Higgs boson in a certain way. In principle, however, new physics will not only change the magnitude of the couplings but also their tensor structure. These new couplings usually are referred to as {\it anomalous couplings}, and the general statistical method of analysis, to be presented in the next chapter, would be applicable in that case as well.

\section{Statistical approach}\label{sec:statistics}
In this work, we will make use of Bayesian probability theory, in which each proposition is associated with a probability or \emph{plausibility}, defined to lie between 0 and 1. This is the only consistent extension of boolean logic incorporating uncertainty \cite{Jaynes:book,Cox:1946,Loredo:1990}.

In Bayesian inference, the laws of probability are used to infer which underlying hypotheses, assumption, or data model \footnote{Typically, \qu{model} will refer to any assumption from which data can be predicted, and not necessarily a full, realistic physical model.} is preferred by some given set of data. Of interest is \emph{Bayes' theorem}, which can be used to reverse the order of the conditioning, denoted by \qu{$|$},
\be \Pr(A|B) = \frac{\Pr(B|A) \Pr(A)}{\Pr(B)}. \ee
Thus, two different hypotheses or models can be compared using the data $\mbD$, through calculation of the \emph{posterior odds}, given by
\begin{equation}\label{eq:post_ratio} 
\frac{ \Pr(M_i|\mathbf{D})}{\Pr(M_j|\mathbf{D})} =
\frac{\Pr(\mathbf{D}|M_i)}{\Pr(\mathbf{D}|M_j)} \frac{\Pr(M_i)}{\Pr(M_j)}. 
\end{equation}
The \emph{prior odds} $\Pr(M_i)/ \Pr(M_j)$ quantifies how much more plausible one model is than the other \emph{a priori}. This ratio is typically taken equal to unity, which however must be considered more carefully in some cases. The \emph{evidence} $\ev_i = \Pr(\mathbf{D}|M_i)$ is the likelihood of the model, a measure of how well the model describes, or rather predicted, the data. The \emph{Bayes factor} $B^i_j = \ev_i/\ev_j$ is the ratio of the evidences of the two models and quantifies how much better $M_i$ describes the data than $M_j$.

Given that the model $M$ contains the free parameters $\mbTh$, the evidence is given by
\bea
\mathcal{Z} =\Pr(\mathbf{D}|M) &=&  \int \Pr(\mathbf{D},\mbTh|M)\df^N\mathbf{\Theta}\notag \\ &=& \int \Pr(\mathbf{D}|\mbTh, M) \Pr(\mathbf{\Theta}|M)\df^N\mathbf{\Theta} 
\notag \\ &=&\int{\mathcal{L}(\mathbf{\Theta})\pi(\mathbf{\Theta})}\df^N\mathbf{\Theta},
\label{eq:Z}
\eea
where $\mathcal{L}(\mathbf{\Theta}) \equiv \Pr(\mathbf{D}|\mathbf{\Theta}, M)$ is the \emph{likelihood function}.  The prior probability density of the parameters is given by $\pi(\mathbf{\Theta}) \equiv \Pr(\mathbf{\Theta}|M)$, and should always be normalized, \ie, it should integrate to unity. The assignment of priors are probably the most discussed and controversial part of Bayesian inference. This is often far from trivial, but nevertheless this assignment is an important, even essential, part of any Bayesian analysis.

The Bayes factors, or rather the posterior odds, are interpreted or \qu{translated} into ordinary language using the so-called \emph{Jeffreys scale}, given in \tabref{tab:Jeffreys} as used in, \eg, Refs.~\cite{Trotta:2008qt,Hobson:2010book} (\qu{$\log$} denotes the natural logarithm). Even though the Bayes factor in general will favour the correct model once \qu{enough} data has been obtained, the evidence is often highly dependent on the choice of prior.

\begin{table}
\centering
\begin{tabular}{|c|c|c|c|}
\hline
$|\log(\text{odds})|$ & odds & $\Pr(M_1 | \mathbf{D})$ & Strength of evidence\\ 
\hline
$<1.0$ & $\lesssim 3:1$ & $\lesssim 0.75$ & Inconclusive \\
$1.0$ & $\simeq 3:1$ &  $\simeq 0.75$ & Weak evidence \\
$2.5$ & $\simeq 12:1$ & $\simeq 0.92$ & Moderate evidence \\
$5.0$ & $\simeq 150:1$ & $ \simeq 0.993$ & Strong evidence \\
\hline
\end{tabular}
\caption{\it The Jeffreys scale, which is used for interpretation of Bayes factors, odds, and model probabilities. The posterior model probabilities for the preferred model are calculated assuming only two competing hypotheses and equal prior probabilities. Note that $\log$ denotes natural logarithm.}
\label{tab:Jeffreys}
\end{table}

Under the assumption that a model $M$ is true, complete inference of its parameters is given by the posterior distribution,
\begin{equation} \label{eq:Bayes_params} 
\Pr( \mathbf{ \Theta} | \mathbf{D},M) = \frac{\Pr(\mathbf{D}
|\mathbf{\Theta},M)\Pr(\mathbf{\Theta}|M)}
{\Pr(\mathbf{D}|M)}  = \frac{\lhood(\mbTh)\pi(\mbTh)}{\ev}.
\end{equation}
In this case, the evidence is only a normalization factor, since it is independent of the values of the parameters $\mbTh$, and it is therefore often disregarded in parameter estimation. However, the actual values of the parameter within a pre-specified model are often not of the greatest interest. Instead, the primary question is usually which model, or set of models, is preferred by the data. 

After model comparison, there might still be a significant amount of uncertainty regarding which model actually is the best, and this uncertainty should not be ignored when making inference on parameters. Model uncertainty can be taken into account by calculating the model-averaged posterior distribution \cite{hoeting1999,Parkinson:2013hzz}
\be  \Pr(\eta | \mbD) = \sum_{i} \Pr(\eta |H_i, \mbD) \Pr(H_i |\mbD), \ee
which is the average of the individual distributions over the full space of the models considered, weighted by the posterior model probabilities. Averaging over models can be done for both prior and posterior distributions, however, the parameters $\eta$, which could be derived, obviously need to be well-defined in all of the models. The posterior in \eref{eq:Bayes_params} is obtained by setting all prior model probabilities, except one, equal to zero. For applications in physics and cosmology, see Refs.~\cite{Parkinson:2013hzz,Vardanyan:2009ft,Vardanyan:2011in}.

The main result of Bayesian parameter inference is the posterior and its marginalised versions (usually in one or two dimensions). Commonly, point estimates such as the posterior mean or median are given together with \emph{credible intervals (regions)}, which are defined as intervals (regions) containing a certain amount of posterior probability. These regions are not unique, without further restrictions, similarly to classical confidence intervals, and in general they do not describe all the information contained in the posterior. We use \MN~\cite{Feroz:2007kg,Feroz:2008xx,Feroz:2013hea} for the evaluation of all evidences and posterior distributions in this work.

\subsection{Model comparison and Higgs couplings}\label{subsec:modelc}

We want to determine whether there is any evidence in the LHC data for deviations from the SM values of the couplings, \ie, if $\kappa_i \neq \kappa^\SM_i$, or if $\kappa_i = \kappa^\SM_i$ is sufficient to describe the data. In other words, we are interested in {\it if} there is a deviation from the SM couplings, and not precisely how large it is, given that it is non-zero. For each coupling this gives two distinct cases and in order to differentiate between them, we want to perform Bayesian model comparison. Note that, from a statistical viewpoint, a model with $\kappa_i = \kappa^\SM_i$ can also be interpreted as a model where there is some non-zero, but negligible (given current data) deviation from the SM value, see Ref.~\cite{hoeting1999} for further discussion.  Beforehand it is not specified whether the other couplings, \ie, the couplings with indices $j \neq i$, should be fixed to their SM value or not, which gives rise to a complication. In principle, there is an important distinction since, without making the assumption of a particular model, any combination of the couplings can deviate from the SM values. 

Thus, we can consider the models $H_\alpha$, with $\alpha = (\alpha_1,\alpha_2, \ldots , \alpha_n)$, where each $\alpha_i = 0$ if $\kappa_i = \kappa^\SM_i$ and $\alpha_i = 1$ if $\kappa_i \neq \kappa^\SM_i$. In total there are $2^n$ models, where $n$ is the number of free parameters. In fact we can consider $\alpha$ as a discrete parameter, for which the posterior odds is given by
\be \label{eq:alphaBayes} \frac{\Pr(\alpha|\mbD)}{\Pr(\beta|\mbD)} = \frac{\Pr(\mbD|\alpha)}{\Pr(\mbD|\beta)}\frac{\Pr(\alpha)}{\Pr(\beta)} = \frac{\ev_\alpha}{\ev_\beta}\frac{\pi_\alpha}{\pi_\beta},\ee
where the calculable Bayes factor $\mathcal{B}^\alpha_\beta=\ev_\alpha/\ev_\beta$ quantifies how much better $\alpha$ describes the data than $\beta$. The natural baseline model is $ \SM = \bar{0} = (0,0,\ldots,0) $, and all the $\mathcal{B}^\alpha_\beta$ can be obtained from the Bayes factors with respect to the SM, $\mathcal{B}^\alpha_\SM$, as $\mathcal{B}^\alpha_\beta = \mathcal{B}^\alpha_\SM / \mathcal{B}^\beta_\SM$. If also finite prior probabilities are assigned to the full set of models, finite posteriors $\Pr(\alpha|\mbD)$ can be calculated, even though we will typically refrain from doing this.
Calculating the Bayes factor does, however, require assignment of priors on the couplings in all the models, which is non-trivial and will be discussed in detail in Sec.~\ref{sec:prior}.

A different, but equivalent, approach is to instead consider a \emph{single} model with a prior which is a mixture of the continuous prior and a point mass at the SM value,
\bea\label{eq:mixedprior} \pi(\kappa_i) &=&    (1-p_{i}) f_i(\kappa_i) + p_{i} \delta(\kappa_i - \kappa_i^\SM), \eea
for each coupling.\footnote{In the general, non-separable case all the quantities in the equation can depend on couplings $\kappa_j$ for $j<i$.} Here the continuous part of the prior, given by $f_i$ (which is normalized to unity), corresponds to the prior assuming $\alpha_i = 1$ and is assigned a total probability $1-p_{i}$, while the SM value of the coupling is assigned a probability $p_{i}$. Note that $\alpha$ is a function of $ \kappa$ and hence that the priors and posteriors of $\alpha$ can be calculated from the distributions obtained using \eqref{eq:mixedprior}. In addition, the Bayes factors (which are independent of the prior on $\alpha$) can be calculated using \eqref{eq:alphaBayes} by factoring out the prior odds. 

\subsection{Inclusion of individual couplings}\label{sec:indcouplings}
In the previous section, we discussed the comparison of $2^n$ models, with different numbers of Higgs scale couplings kept free. However, when $n$ grows in size, comparing this large number of models to each other rapidly becomes less transparent. 

One can test if a particular variable should be included by comparing the cases $\kappa_i = \kappa_i^{\SM}$ and $\kappa_i \neq \kappa_i^{\SM}$, and hence calculating the Bayes factors
\be \mathcal{B}_i =\frac{\Pr(\mbD|\alpha_i = 0)}{\Pr(\mbD|\alpha_i = 1)}. \label{eq:BF_single}\ee
Again, however, one has to decide what to do with the \emph{other} couplings, \ie, which priors to assign them. Possibilities could be
\begin{enumerate}[(i)]
\item fixed to the SM value $(\mathcal{S})$, 
\item free and different from the SM value $(\mathcal{F})$, or
\item either of the above, \ie, an average $(\mathcal{A})$.
\end{enumerate}    
The evidences are given by the likelihoods integrated not only over the prior on $\kappa_i$, but also over the prior on all other couplings. In particular, 
\be \label{eq:single_av}  \Pr(\mbD|\alpha_i) = \sum_{\alpha_i^*} \Pr(\mbD|\alpha_i, \alpha_i^*)\pi(\alpha_i^*),\ee which depends on the prior on $\alpha_i^* = (\alpha_1,\ldots, \alpha_{i-1},\alpha_{i+1},\ldots, \alpha_n)$. The evidences in Eq.~\eqref{eq:single_av} are simply the evidences discussed in Sec.~\ref{subsec:modelc}. The three cases then corresponds to $p_k = \pi(\alpha_k =0)$ being equal to either $p_k = 1, 0$, or some intermediate values, most naturally $0.5$ (see \cite{hoeting1999} for detailed discussion). The results are only expected to be independent of this choice in the case where the constraints on one parameter is independent of the values of the others.

\subsection{Single comparison with SM}\label{subsec:single}
In physics there is often a theoretically \emph{a priori} motivated \qu{baseline} model which all extended models are usually compared to. In the present case the obvious choice for such a reference model is the SM. Furthermore, Bayesian model comparison treats all models on equal footing, which enables quantification of how much the SM is favoured with respect to extended models. Again, this could be done in the context of specific renormalizable high-energy models, but here we will focus on the effective case only considering the rescaled couplings.

We want to compare the SM with a model \qu{not-SM}, or $\nSM$. The question is how this model for comparison should be defined. For example, one could compare the SM with a model with only a single coupling free, which is just one of the cases discussed in the previous chapter. However, this is obviously not satisfactory since there are many such models\footnote{This could be remedied by comparing with a model in which any of the couplings are free, but only one at a time.}, and at the same time we are completely neglecting models with two or more couplings free\footnote{Again, the constraint $\kappa_i = \kappa_i^\SM $ can also be interpreted as holding only to a very good approximation, but not exactly.}.
Alternatively, one could compare with the most general model in which all couplings are free. However, the issue is the same, still neglecting the possibility that there could be significant deviations in more than one coupling, but not in all at once. The most general model could be punished for the inclusion of the couplings for which the SM value is preferred. Therefore, the most appropriate comparison appears to be the one between the SM and a model in which each coupling \emph{either} takes the SM values, or differs from it.
 
Indeed, probability theory again yields
\be \Pr(\mbD|\nSM) = \sum_{\alpha} \Pr(\mbD|\alpha) \pi(\alpha|\nSM), \ee
and all the above cases are just cases for a specific choice of prior $\pi(\alpha|\nSM)$. Due to lack of further information, we take $\pi(\alpha_i|\nSM) = 1/2 $, which means that in the $\nSM$ model it is equally probable that each coupling deviates (significantly) from the SM value, as it is that there is no (or negligible) deviation. In this case, however, the couplings of the model $\nSM$ equal the SM couplings with prior probability $1/2^n$. This part of $\nSM$, \ie, the part that is statistically equivalent to the SM, can of course then just be excluded in the analysis, and this will be our default choice.
In principle, however, one could also motivate its inclusion by saying there could still be a deviation from the SM values, but a negligible one. Note that adding any additional couplings, which are unconstrained by data, does not affect the comparison of SM and $\nSM$.

\subsection{Choice of prior}\label{sec:prior}

As discussed in Sec.~\ref{sec:higgs}, we will consider the two cases: (i) all tree-level couplings are allowed to vary, with the loop-induced couplings calculated assuming no additional contribution from new physics; (ii) all couplings, including the loop-induced ones, are allowed to vary, which implies that new physics is allowed to participate in the loop processes. In the first case there are 6 free parameters $(\kappa_t,\kappa_b,\kappa_\tau,\kappa_W,\kappa_Z,\kappa_\mu)$, whereas there are 8 free parameters in the second case, adding ($\kappa_\gamma,\kappa_g$). Note that the default \qu{SM} values of these couplings are those calculated in Eqs.~\eqref{eq:gluon} and \eqref{eq:gamma} assuming no new particles, which do not necessarily correspond to the actual exact SM value (equal to 1). In addition, the value of the scale factor for the total Higgs width, $\kappa_H$, will depend on the other ones according to Eq.~\eqref{eq:kappah}. We shall however not consider this as a free parameter in either case.

In order to calculate the evidence of the models in which the couplings differ from the SM value, a prior for each coupling $\kappa_i$ is needed. The assignment of prior is an important task since not only the posteriors within each model depend on it, but perhaps more importantly, so does the evidence. It is therefore important to take care to include as much known information into the prior without making any assumptions based on the data under consideration.
\begin{itemize}
\item \emph{Default: uniform.} A common choice is to take a uniform prior on each of the couplings in order to implement \emph{a priori} \qu{ignorance}, usually unbounded or with \qu{wide enough} limits. However, such a uniform prior cannot quantify ignorance in a parameter, if not only because a uniform prior in one parameter will not be uniform in a parameter given by a non-linear transformation of the first one. 
Secondly, an unbounded (improper) prior often gives meaningless answers for the evidence, and so do many priors in the limit when their widths go to infinity. However, this does not necessarily imply that the uniform prior as such is useless or in general should be avoided. As any prior it can be used when it is motivated, and we shall use it in \secref{sec:defparam} to get a rough idea of what the parameter constraints on the different couplings are.

\item \emph{Couplings: uniform.}
In the case where only the tree-level couplings are free, one can consider the actual couplings appearing in the Lagrangian as the free parameters. In the Higgs sector there are Yukawa couplings for the Higgs coupling to fermions as well as the Higgs coupling to the gauge bosons. In principle, one could argue that \emph{a priori} all couplings should be of order one. Hence, a roughly uniform prior on each of the couplings, with an effective upper limit of some constant of order one would seem appropriate.\footnote{See \refcite{Bergstrom:2014owa} for a similar application in the lepton sector of the SM and Ref.~\cite{Fichet:2012sn} for a more general discussion of naturalness in Bayesian inference.} However, if the measured couplings have a small (absolute) errors compared to one, this will lead to a very strong \qu{Occam effect}
which will strongly disfavour modifications of the couplings and give strong preference to the SM values.
This is indeed the case, with the possible exception of the top Yukawa, and the masses of the SM particles differ by many order of magnitudes, a discrepancy commonly known as the \qu{flavour puzzle} \cite{Libanov:2011st}.
 Hence, all models with additional couplings will be severely disfavoured if this prior is taken, and so we will not perform a detailed analysis of this case, even though these conclusions are worth to bear in mind.

\item \emph{Logarithmic.} Dropping the assumption that the couplings should be of order one, it might seem more appropriate that instead the order of magnitude of the couplings are \emph{a priori} unknown. Thus, the choice is instead a logarithmically uniform prior on $y_i$ between some lower limit and the perturbative upper limit, taken as $4\pi$. The lower limit must be chosen by hand; we will use $10^{-7}$ as the default choice. However, it turns out that the results are insensitive to changing this lower limit by at least a few orders of magnitude. Furthermore, for simplicity we will always assume positive couplings. In most cases the sensitivity to the sign of the couplings is small, with the exception of the case when $\kappa_W$ and $\kappa_t$ have different signs, which can enhance the rate of the $H\rightarrow \gamma\gamma$ \cite{CMS-PAS-HIG-14-009, ATLAS-CONF-2014-009}. However, there is no clear sign of this enhancement in the data, which implies that the total mass of the mode in that region will not be much larger than in the region with positive couplings, and therefore the effect on the evidences will be very small.

\item \emph{Gaussian.} Instead of assuming, as in the previous cases, that the couplings are \emph{a priori} unrelated to the SM ones, one can consider that many SM extensions, such as the ones mentioned in Sec.~\ref{sec:higgs}, will all lead to modifications of roughly the same size as the SM couplings. Without considering a specific model, we cannot determine the sizes of these contributions, nor if they should be positive or negative. Hence, all we can do is to say that we \emph{a priori} expect $\mean{\kappa_i} = \kappa_i^\SM$, and a typical deviation of $\sigma(\kappa_i) = s_i = \mathcal{O}(1)$. Out of all the (prior) distributions on the real numbers with these constraints there is a unique one which has maximal entropy (or equivalently \qu{minimal information}), namely the Gaussian distribution \cite{Sivia:1996,Jaynes:book}. We will consider values $s_i = s$ in the range $1-4$ as appropriate, with a default value of $s=2$.
\end{itemize}

Finally, we mention that one in principle could consider the SM augmented with additional higher-dimensional effective operators. These modify the SM couplings by an amount proportional to ${v^2}/{\Lambda^2}$, where $\Lambda$ is the scale of new physics \cite{Corbett:2012ja,Corbett:2012dm,Contino:2013kra, Fichet:2013jla,Dumont:2013wma}. These operators could be implemented in a Bayesian analysis such as in \refcite{Dumont:2013wma, deBlas:2014ula}, but one could also utilize the expected sizes of the additional contributions in the present analysis by using a prior on $\Lambda$ and from this obtain priors on the $\kappa_i$'s. If one expects that $\Lambda$ could be of any order of magnitude, much of the prior would be piled up close to the SM values, which would imply that it would be possible to obtain significant evidence against the couplings taking those values, but not in favour. However, if the scale of new physics is assumed to be close to the electroweak scale as in \refcite{Dumont:2013wma}, the typical modification would be of order one, in which case one will get a result similar to the one for the Gaussian prior above.

To summarize, in the present work we shall consider the following models and priors. In the case with only the tree-level couplings free we shall make the analysis both using a logarithmic prior, which is placed directly on the actual couplings, and a Gaussian prior, which instead is placed on the coupling scale factors, $\kappa_i$. In the second case, where both tree and loop-level couplings are free, we shall only make an analysis using the Gaussian prior placed on the scale factors. In this case it should be noted that the expectation value of the now free parameters $\kappa_g$ and $\kappa_\gamma$ are the values given from the other scale factors, \ie, the values given by Eqs.~\eqref{eq:gluon} and \eqref{eq:gamma}, and not the SM value of these scale factors (which is 1).

\section{Results}\label{sec:results}

\subsection{Default parameter constraints}\label{sec:defparam}
In this section, we obtain the \qu{default} parameter constraints on the coupling scale factors by calculating the likelihood using \HS~and imposing a uniform prior on the $\kappa$'s with zero as the lower limit and a \qu{large enough} upper limit. Although this prior does not impose \emph{a priori} ignorance, and it cannot be used  for model comparison, the derived parameter constraints will be valid as long as the uniform prior is reasonable in the the region of parameter space which are not completely ruled out by the data.  A fixed Higgs boson mass of $m_H=125.5$ GeV was used, and will be used throughout this work. 

Similar to the model comparison performed later, we first simultaneously estimate only the scale factors present at tree-level, and then additionally also the loop-induced scale factors. In addition to these two cases, we shall consider the special case where new physics only contribute to the loop-induced processes and thus only the scale factors corresponding to these processes, \ie, $\kappa_g$ and $\kappa_\gamma$, are free.

\begin{center}
\begin{figure}
\includegraphics[width=\textwidth]{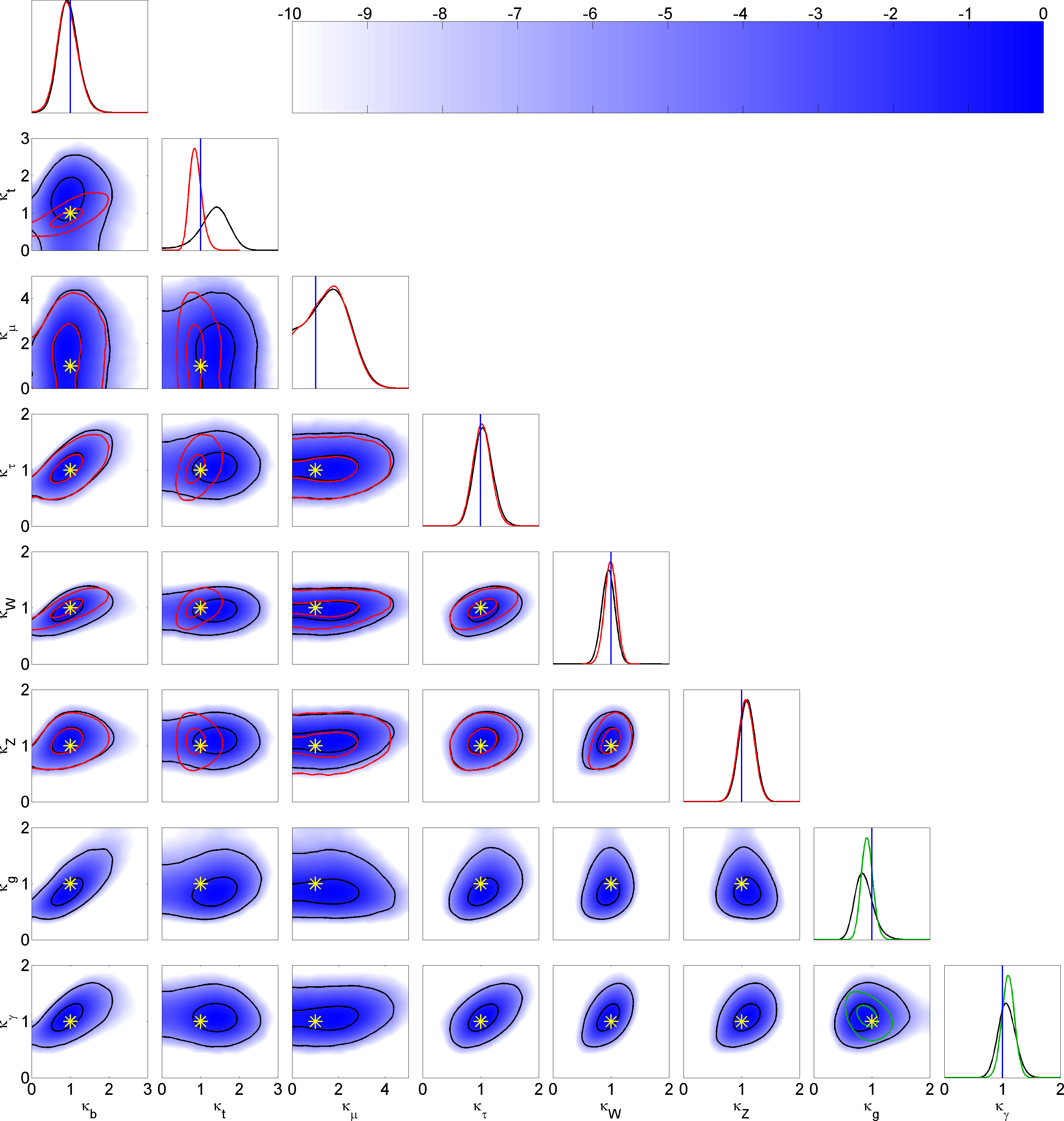}
\caption{Results of (default) parameter estimation. Two-dimensional log-posterior distribution (blue shading), $1\sigma$ and $3\sigma$ Bayesian credible regions and one-dimensional posteriors (both black) of the eight-parameter fit.
$1\sigma$ and $3\sigma$ credible regions and one-dimensional posteriors of the tree-level six-parameter fit (red) and the two-parameter fit (green). The SM values of unity are marked with vertical lines and stars, respectively.}\label{fig:de_param_tri}
\end{figure}
\end{center}

In Fig.~\ref{fig:de_param_tri} we present the results in terms of one- and two-dimensional posterior distributions. In the two-dimensional plots the blue shading denotes the natural logarithm of the posteriors and the black contours the $1\sigma$ and $3\sigma$ credible regions, \footnote{Defining the contours by the usual $\chi^2$-thresholds on $ - 2 \log(\lhood(\theta)/\lhood_{\rm max})$, with $\lhood(\theta)$ the Bayesian marginal likelihood, yields virtually identical contours in all cases.} while the one-dimensional posteriors are also black in the plots on the diagonal. 
Superimposed on these, in red, are the $1\sigma$ and $3\sigma$ contours as well as the one-dimensional posteriors for the case when only the tree-level scale factors are free. As previously discussed, $\kappa_g$ and $\kappa_\gamma$ are given as functions of the free scale factors. Finally, the same quantities are presented in green (in the bottom right) for the case when the tree-level couplings remain fixed at their SM values but new physics is allowed to participate in the loop-induced processes. The SM values are marked with stars and vertical lines.

In all three fits, \emph{all} the SM values are inside (or extremely close to) the $1\sigma$ regions, which is in fact rather unlikely.
As expected, adding $\kappa_g$ and $\kappa_\gamma$ to the set of free parameters will relax constraints on the six free tree-level couplings. The main effects should be seen in the scale factors corresponding to the particles which give the main contribution to the loop processes. Thus, the largest effect will be for the top quark which gives the absolutely dominating effect to the loop in the gluon fusion process, while a smaller effect should also be seen in the bottom quark coupling. Apart from the top quark this is also the only particle that participates in both the gluon fusion and $H\rightarrow \gamma\gamma$ processes. The modifications to the other couplings are marginal. In a similar manner, the constraints on the loop-induced couplings are weaker in the eight-dimensional fit than in the two-dimensional one.

Finally, from the plots in Fig. 1 we can conclude that there is quite strong support for the couplings having non-zero values, with the exception of $\kappa_\mu$, $\kappa_t$, and to some extent $\kappa_b$, in the eight-parameter fit.

\subsection{Model comparison: all models}

Although the previous results were interesting, they were all derived under the assumption that the scale factors actually differed from those of the SM. Following \secref{sec:statistics} we would instead like to perform model comparison. We will use the priors discussed in \secref{sec:prior} and aim to evaluate how much the results depend on these different prior choices.

In this section we follow \secref{subsec:modelc} and compare models with any combination of free parameters. In particular, we use \MN~with the priors in \eqref{eq:mixedprior} and $p_i=p_0$ chosen so that the posterior over the space of models becomes as uniform as possible, and so all values of $\alpha$ will be sampled adequately \footnote{We check that the statistical error on each Bayes factor is reasonably small by considering the effective sample size $(\sum w_i)^2/\sum w_i^2$, with the $w_i$'s the weight of each sample belonging to a certain model. Typically it is of the order of $10^2 - 10^3$, but in a few cases slightly smaller.}. There are in total $2^n$ models, with $n=6$ when the tree-level couplings are free and $n=8$ when also loop-processes are included. 

\begin{figure}
\begin{center}
\includegraphics[width=0.495\textwidth]{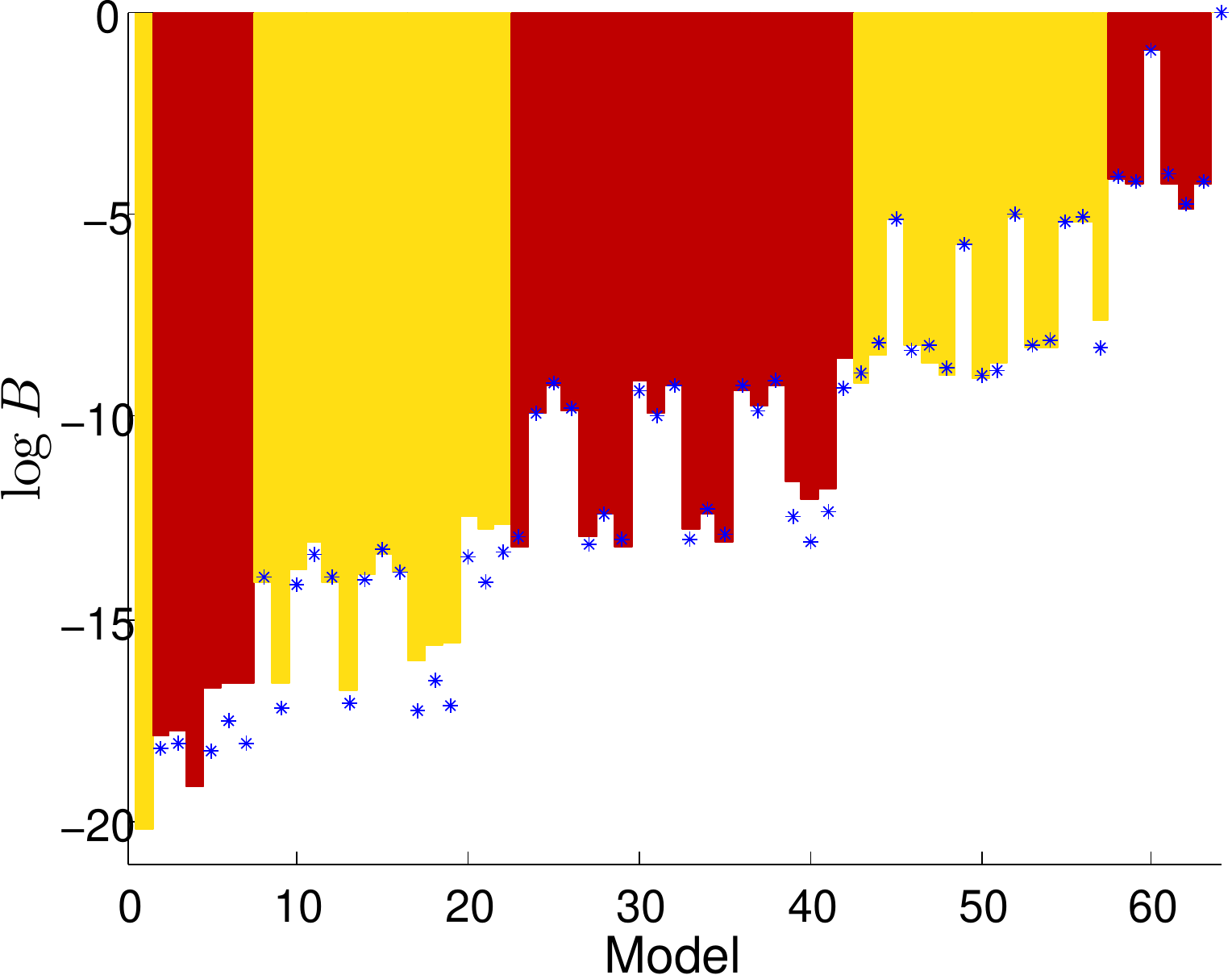}
\includegraphics[width=0.495\textwidth]{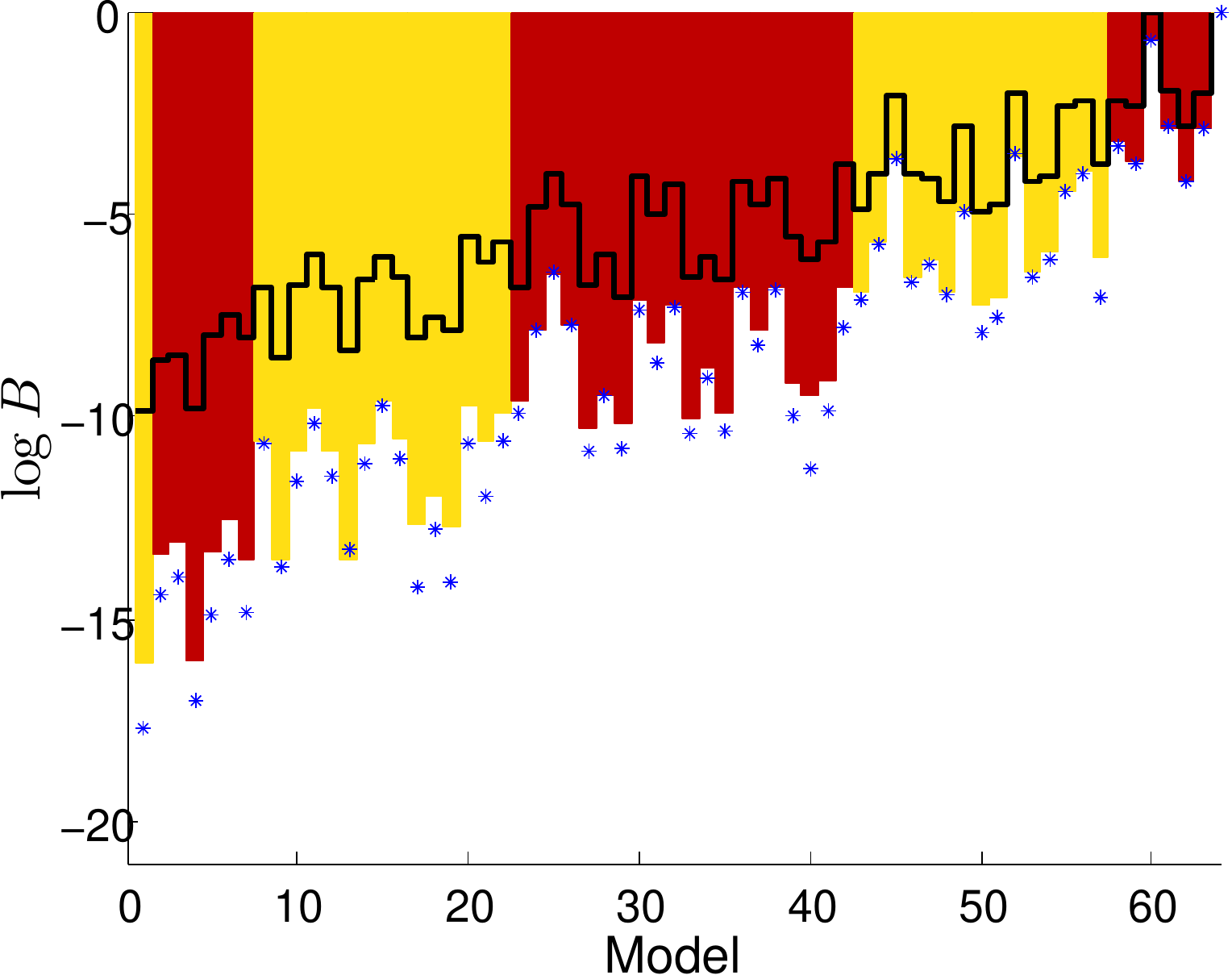}
\caption{Left: logarithms of Bayes factors (with respect to the SM) with the logarithmic prior on tree-level couplings. Each unicoloured block of bars have the same number of free parameters, from the right: 0 (the SM), 1, etc., to the most general model with all couplings free to the far left. The blue stars are the values obtained by extrapolating the the values of the single-coupling models, assuming independence. Right: same as the left plot but with Gaussian priors. The bars correspond to a standard deviation of $s=4$ and the solid black line to $s=1$.} \label{fig:logB6}
\end{center}
\end{figure}

In the left panel of \figref{fig:logB6} we present the logarithms of Bayes factors for all of the $2^6=64$ models, compared to the SM and using the logarithmic prior (on $[ 10^{-7},4\pi]$) for the tree-level couplings, \ie, the Yukawa couplings and gauge boson couplings. The models are divided into unicoloured groups depending on the number of couplings which are free. In the model to the far left in the figure all couplings are free, the models in the next group have 5 parameter free, etc., until the model to the far right, which is the SM (and has no visible bar since $\log B = 0$). The blue stars are the values calculated by extrapolating the comparison of the SM with the models with a single coupling free, and then assuming that adding an additional parameter has the same effect on $\log B$ regardless of the assumptions on the other parameters. This would be exact if the shape of the likelihood as a function of each parameter did not depend on the values of the other parameters. Although not exact, it seems that treating all of the parameters as independent gives a reasonable approximation for the model comparison.

As expected, there is a clear trend. The larger the number of free couplings, the smaller the values of $\log B$, \ie, the stronger the evidence against that model. Hence the evidence against the model with all couplings free is very strong. Adding any of the parameters makes the model worse with about the same amount, with the exception of $\kappa_\mu$, which only decreases the evidence with a small amount (roughly one log-unit). Letting $\kappa_W$ free, corresponding to most heavily constrained coupling, will have the largest effect on the evidence of the model. 

Furthermore, one should remember that the log-odds only equals $\log B$ when the priors are equal. In this case, one might argue that the SM should have a larger prior than any of the other models, perhaps the same as all the other models together, which (assuming that prior is uniformly distributed) would lead to the log-odds being $\log 2^6 \simeq 4$ smaller than the $\log B$'s in the plot. Again, we note that the dependence on the prior limits is very weak. For example, decreasing the lower limit to $10^{-15}$ would lead to a decrease of $\log B$ smaller than $0.7$ for the addition of each coupling.

In the right panel of \figref{fig:logB6} we present $\log B$ for the same models, but with Gaussian priors on the coupling scale factors. The bars are obtained using a standard deviation of $s=4$, and the solid black line using $s = 1$.  Naturally, the choice of priors affects the exact values of the evidences, but the general trend is the same in all cases. Adding a parameter with a Gaussian prior is not as influential as adding one with a logarithmic prior, and the difference between the two Gaussian priors is only about one log-unit per parameter. 

Next, we consider the case when also the loop-induced couplings are allowed to differ from the SM values, or rather those calculated in Eqs.~\eqref{eq:gluon} and \eqref{eq:gamma}, giving a total of $2^8=256$ combinations of free couplings. The same Gaussian priors as in the right panel of \figref{fig:logB6} has been used, but with the expectation values of $\kappa_g$ and $\kappa_\gamma$ given by Eqs.~\eqref{eq:gluon} and \eqref{eq:gamma}, since this is the expectation without any contribution from new physics. The trend is similar to the previous case with tree-level couplings in that models with few free couplings are preferred to models with more free couplings. However, when approaching the models with most free parameters, there seems to be a \qu{levelling off} in the sense that adding more parameters is less damaging. This makes sense because, if the parameter constraints deteriorates when more free parameters are added, the evidence will tend to be larger than what would otherwise be expected. Finally, in a similar way to the previous case on one might consider the SM not on equal footing with each of the other models, making the posterior odds smaller than the Bayes factor (now with $\log 2^8 \simeq 5.5$ log-units).

\begin{center}
\begin{figure}
\includegraphics[width=0.9\textwidth]{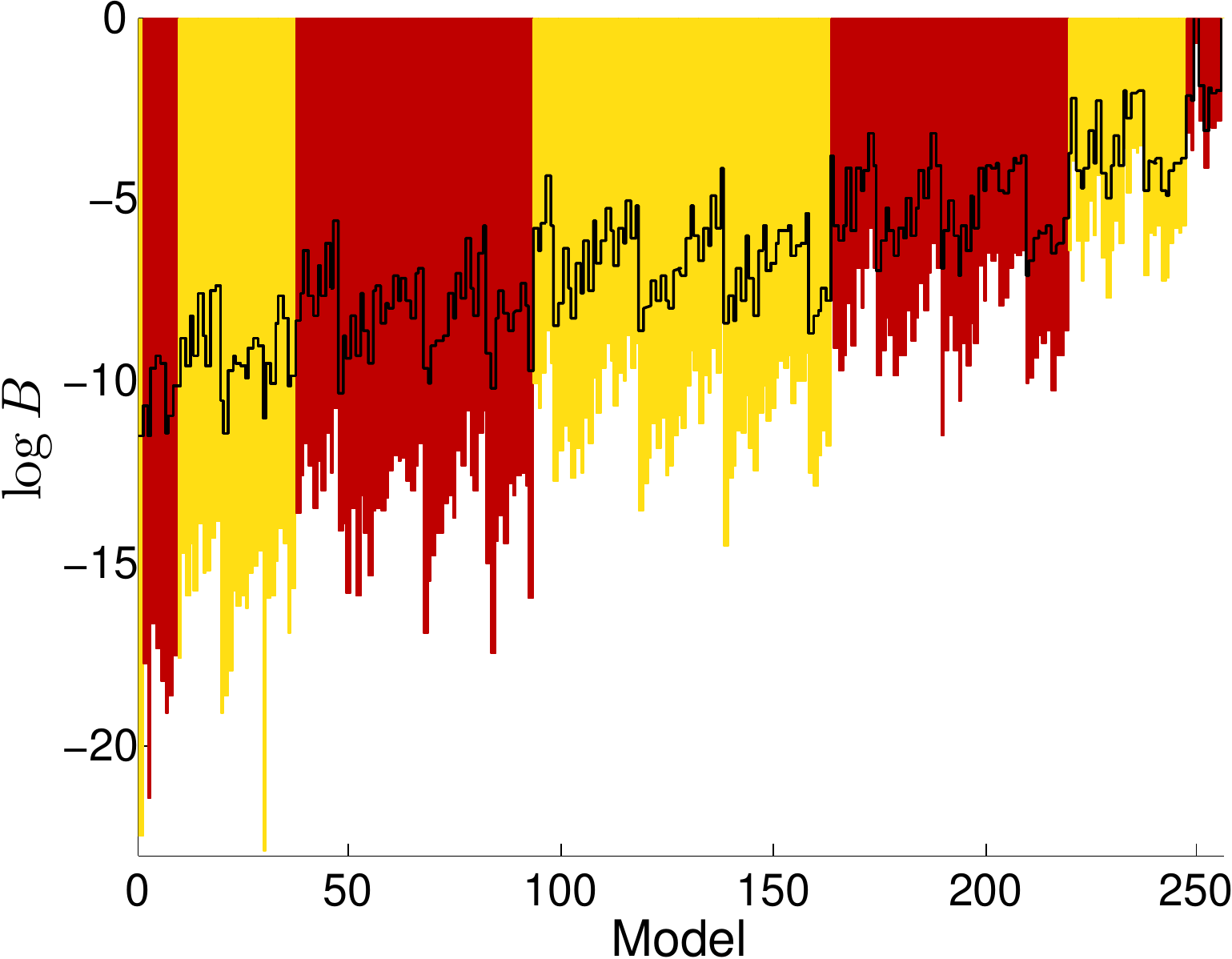}
\caption{Logarithms of Bayes factors (with respect to the SM) for models with up to 8 free parameters, with a Gaussian prior with standard deviations $s=4$ (bars) and $s=1$ (black line).} \label{fig:gauss8}
\end{figure}
\end{center}

\subsection{Inclusion of individual couplings}
In the previous section we studied how all the different combinations of free couplings compared to each other. Although some conclusions could be drawn, the result was not completely transparent. In this section we instead follow \secref{sec:indcouplings} and evaluate the evidence for or against the inclusion of each individual coupling.

\begin{figure}
\begin{center}
\includegraphics[scale=0.7]{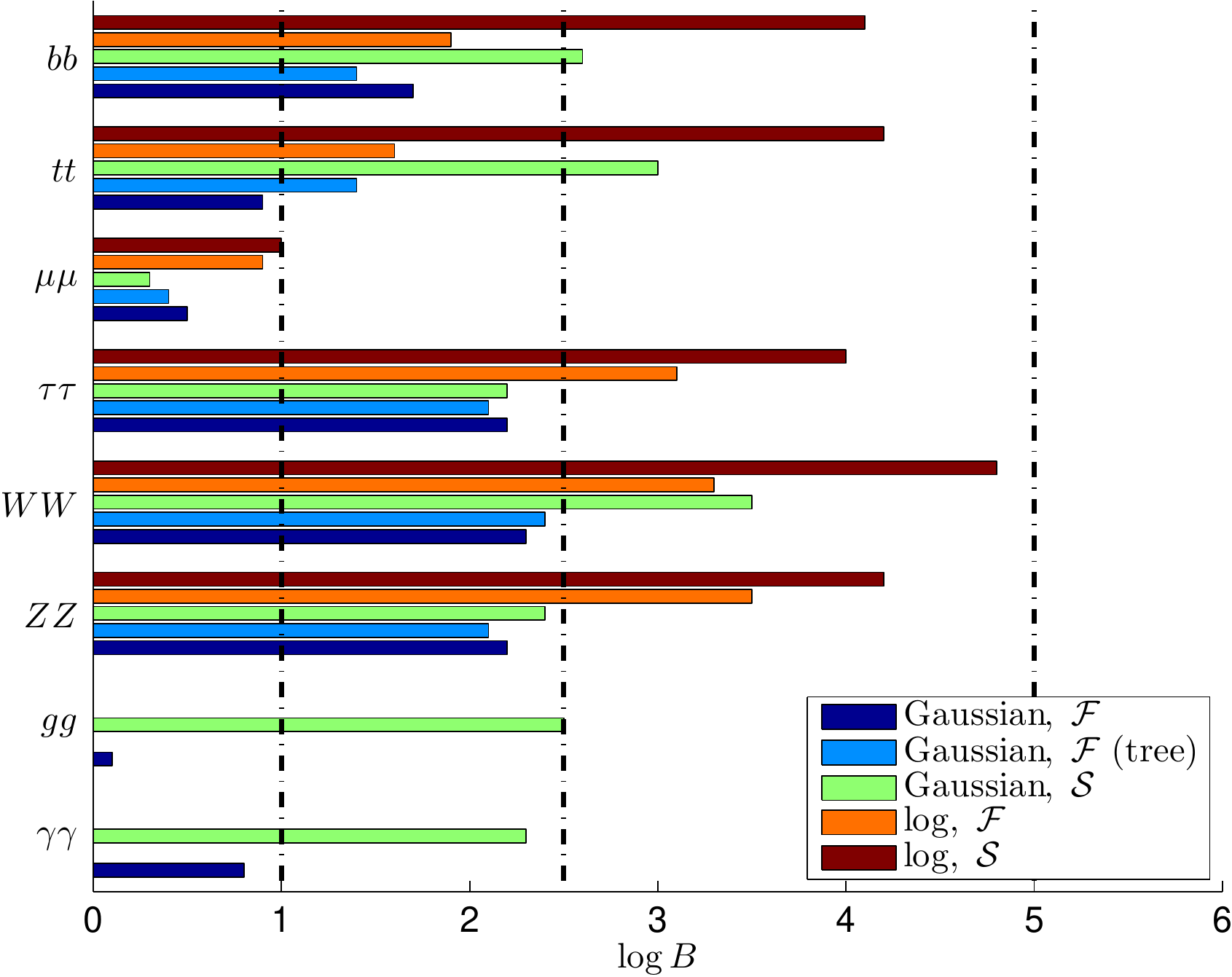}
\caption{ Logarithms of Bayes factors against inclusion of couplings for the eight coupling scale factors. Values larger than $0$ means the SM value of the coupling is prefered. The other couplings are either fixed to their SM values ($\mathcal{S}$), or allowed to vary with the same prior as the coupling of interest ($\mathcal{F}$), or averaged over these two cases ($\mathcal{A}$). Since typically the evidences are much larger when the other couplings equal their SM values, the average is dominated by these components, and hence $\mathcal{A}$ yields essentially the same result as $\mathcal{S}$. \label{fig:logB_coup} }
\end{center}
\end{figure}

In \figref{fig:logB_coup} we give the logarithms of the Bayes factors in \eref{eq:BF_single}, \ie, \emph{against} the inclusion of each of the couplings, both for the case of the six tree-level couplings with logarithmic priors, and for the Gaussian priors on the scale factors. Here we use the value $s=2$ for the standard deviation, although the difference from $s=4$ and $s=1$ as used previously is expected to be quite small. As in \eref{eq:single_av}, the other (nuisance) couplings are either fixed to their SM values ($\mathcal{S}$), allowed to vary with the same priors as the coupling of interest ($\mathcal{F}$), or averaged over these two cases ($\mathcal{A}$). However, in \eref{eq:single_av}, the size of each contribution is proportional to the evidence of that particular model, and since typically the evidences are much larger when the other couplings equal their SM values \footnote{The muon coupling is an exception, but since this is due to a lack of constraints rather than the existence of a tension with the SM value, this has no effect.}, the average is dominated by these components. Hence the result for $\mathcal{A}$ equals that of $\mathcal{S}$ to a very good approximation. Note that these Bayes factors are evaluated separately using dedicated \MN~runs. Hence, these number might differ somewhat from those which can be read from Figs.~\ref{fig:logB6} and \ref{fig:gauss8}. The Bayes factors in the table have significantly smaller numerical errors of about $0.1$.

Some general conclusions which can be drawn are that the logarithmic prior yields a stronger preference for the SM couplings than the Gaussian (as in previous chapter), and $\mathcal{S}$ stronger than $\mathcal{F}$ (which is reasonable since the constraints are relaxed). 

The Higgs decay to $\mu^+\mu^-$ is rather weakly constrained and the results for this coupling is quantitatively different to the other tree-level couplings. For the logarithmic priors there is barely weak evidence in favour of the SM, while for the Gaussian case there is not even that. Moving on to the other tree-level couplings, for the log prior there is weak to moderate evidence for all the couplings, with $\mathcal{F}$ giving about $1-2$ log-units weaker preference than $\mathcal{S}$ and $\mathcal{A}$. For the Gaussian prior, the evidence is also weak to moderate, but typically weaker than the logarithmic case. For the Gaussian prior for the tree-level couplings, there is no significant difference between the cases where the loop-induced couplings are free or not. 

The loop-induced couplings enter only in two cases, both with a Gaussian prior. When the tree-level coupling scale factors are fixed, there is just moderate evidence in favour of the SM values for both $\kappa_g$ and $\kappa_\gamma$, while in the case when the other couplings are free, this preference essentially disappears completely.

However, as discussed in \secref{sec:indcouplings}, making the weakest assumption on the tree-level couplings, Bayesian probability theory tells us that one really ought to use the model-averaged results (the cases with \qu{$\mathcal{S}$} in \figref{fig:logB_coup}). Hence, we conclude that the couplings moderately prefer the SM values for $b\bar{b} $, $t\bar{t}$, $WW$, ZZ, and $\tau^+\tau^-$ for both logarithmic and Gaussian priors. For gg and $\gamma\gamma$ the preference is barely moderate, and for the coupling to $\mu^+\mu^-$, the evidence is barely weak or none at all. 

\subsection{$\text{SM}$ vs $\nSM$}
We consider the $\nSM$ model as discussed in Sec.~\ref{subsec:single}, with the most appropriate assumption is that all the couplings can either take their SM value, or differ from it, with a prior probability of $0.5$ for each. The special case where all couplings simultaneously take on their SM values would typically be excluded from, but could also be included in, the $\nSM$.

In \tabref{tab:logB_p0}, we present the comparison of $\nSM$  with the above model for these two cases and for the different continuous priors.
When the SM part is excluded, the evidence for the SM is actually just about strong for the logarithmic and Gaussian (with $s = 2$) priors on all couplings, and moderate for the case of tree couplings. In the present case, the evidence of  the $\nSM$ is dominated by the contribution from models with a single coupling free, weighted by their priors within the $\nSM$.

In the second case, there is also a contribution from the part equivalent with the SM, which can be relatively large (and even dominating in the logarithmic case).
Still, the conclusions do not change significantly, although the $1.1$ log-units difference for the logarithmic prior takes the evidence for the SM from just about strong to moderate.

\begin{table}
\begin{center}
\begin{tabular}{| c | c | c | c | c | c | c | c| c | c| }
\hline
 Prior & $\log B$  
  \\ \hline
 Log (with SM) & -3.8    
  \\ \hline
 Log (no SM) &  -4.9 
  \\ \hline
 Gauss (tree, with SM)&  -3.1
  \\ \hline
 Gauss (tree, no SM)&  -3.5  
  \\ \hline
 Gauss (all, with SM) &  -4.4 
  \\ \hline
 Gauss (all, no SM)&   -4.8 
 \\ \hline

\end{tabular} 
\caption{$\log B$ between SM and $\nSM$ for three different priors, in the two cases when the values of the couplings is either included in $\nSM$ or not.
}
\label{tab:logB_p0}

\end{center}
\end{table}

\section{Summary and Conclusions}\label{sec:conclusions} 
We have performed a Bayesian analysis of the LHC Higgs data and used an interim framework where the magnitude of the Higgs couplings are rescaled by coupling scale factors, whereas the tensor structure of the couplings is unaltered with respect to the SM. In the present work, we have limited our discussion to the couplings which are constrained by the LHC, in total six tree-level couplings and two loop-induced couplings.

We have performed Bayesian parameter inference on these coupling scale factors in the following three cases: either the tree-level couplings, the loop-level couplings, or both simultaneously, were free. In each case the SM values were well within the $1\sigma$-region. However, when all couplings were free, neither $\kappa_t$ nor $\kappa_\mu$ were well-constrained and could in principle be zero. 

Since the most important question is rather that of which model best describes the data, we have instead focused on Bayesian model comparison, considering models with either only the tree-level couplings in the Lagrangian, or all couplings, allowed to vary. In the first case, we used both a logarithmic prior, which was imposed directly on the tree-level couplings, and a Gaussian prior, imposed on the coupling scale factors. In the second case, when the loop-induced couplings were also treated as free parameters, the analysis was made with a Gaussian prior imposed on the coupling scale factors. In each case we performed model comparison between models with one, several, or all of the couplings free. The larger the number of free parameters, the more disfavoured the model was. 

We have considered a single coupling at a time in the cases where the other couplings could either be fixed to the SM values or allowed to vary with the same prior as the coupling of interest. The favoured models are those with the couplings fixed to the SM value, although the evidence is virtually non-existent for the coupling to $\mu^+\mu^-$. All this was performed with the combinations of free parameters and priors discussed above. Finally, we discussed the definition of the model $\nSM$, and compared this single model to the SM, finding that the SM is moderately to strongly favoured.


\section*{Acknowledgments}
J.B.~acknowledges partial support from the European Union FP7 ITN INVISIBLES (Marie Curie Actions, PITN- GA-2011- 289442). This work was supported by the Swedish Research Council (Vetenskapsr{\aa}det), contract no. 621-2011-3985 (S.R.)


\clearpage

%
\end{document}